\documentstyle[aps,prd,epsf]{revtex}

\begin{document}

\title{Big Bang Nucleosynthesis with Gaussian Inhomogeneous Neutrino Degeneracy}
\author{Spencer D. Stirling$^{1}$ and Robert J. Scherrer$^{2,3}$}
\address{$^1$Department of Mathematics, University of Texas,
Austin, TX~~78712}
\address{$^2$Department of Physics, The Ohio State University,
Columbus, OH~~43210}
\address{$^3$Department of Astronomy, The Ohio State University,
Columbus, OH~~43210}

\maketitle
%

\renewcommand{\baselinestretch}{1.3}
\begin{abstract}

We consider the effect of inhomogeneous neutrino degeneracy on
Big Bang nucleosynthesis for the case where the distribution
of neutrino chemical potentials is given by a Gaussian.  The chemical
potential fluctuations are taken to be isocurvature, so that
only inhomogeneities in the electron chemical potential are relevant.
Then the final
element abundances are a function only of the baryon-photon
ratio $\eta$, the effective number of additional neutrinos $\Delta N_\nu$,
the mean electron neutrino degeneracy parameter $\bar \xi$, and the rms fluctuation
of the degeneracy parameter,
$\sigma_\xi$.
We find that for fixed $\eta$, $\Delta N_\nu$, and $\bar \xi$,
the abundances of $^4$He, D, and $^7$Li are, in general, increasing
functions of $\sigma_\xi$.  Hence, the effect of adding a Gaussian
distribution for the electron neutrino degeneracy parameter is to {\it decrease} the allowed
range for $\eta$.  We show that this result can be generalized to
a wide variety of distributions for $\xi$.

\end{abstract}
%

\twocolumn

\section{Introduction}

Many modifications to the standard model of Big Bang
nucleosynthesis (BBN) have been explored \cite{malaney}.
One of the most exhaustively
investigated variations on the standard model is neutrino
degeneracy, in which each type of neutrino is allowed
to have a non-zero chemical potential
\cite{wagoner}, and a number of models
have been proposed to produce a large lepton degeneracy
\cite{gelmini}-\cite{murayama}.
More recently, observations of the cosmic microwave background
(CMB) fluctuations have been combined with BBN to further constrain
the neutrino chemical potentials
\cite{lesg}-\cite{orito2}.

An interesting variation on these models is the possibility that
the neutrino
degeneracy is
inhomogeneous \cite{dolgov1}-\cite{dibari}.  The
consequences of inhomogeous neutrino degeneracy for BBN were
examined by Dolgov and Pagel \cite{dolgov2} and
Whitmire and Scherrer \cite{whitmire}.
Dolgov and Pagel examined models in which the length scale of
the inhomogeneity was sufficiently large to produce an inhomogeneity
in the presently-observed abundances of the elements produced in BBN.
Whitmire and Scherrer investigated
inhomogeneities in the neutrino degeneracy
on smaller scales; in these models the element abundances
mix to produce a homogeneous final element distribution.
Using a linear programming technique, they derived  
upper and lower bounds
on the baryon-to-photon ratio $\eta$ for {\it arbitrary}
distributions of the neutrino chemical potentials and showed
that the upper bound on $\eta$ could be considerably relaxed.
However, the resulting distributions for the neutrino chemical potentials
were quite unnatural.  Hence,
in this paper, we examine a
more restricted class of models, in which the distribution of the chemical
potentials is taken to be a Gaussian.

In the next section, we discuss our model for inhomogeneous
neutrino degeneracy.
We calculate the effect of these inhomogeneities on
the final element abundances and discuss our results
in Sec. 3.
We find that, in most cases, the effect of Gaussian
inhomogeneities in the electron
neutrino chemical potential is to increase the abundances of
deuterium, $^4$He, and $^7$Li relative to their abundances
in models with homogeneous neutrino degeneracy.

\section{Model for Inhomogeneous Neutrino Degeneracy}

We first consider the case of homogeneous neutrino degeneracy.
For this case, each type of neutrino is characterized
by a chemical potential $\mu_i$ ($i = e, ~\mu, ~\tau$),
which redshifts as the temperature, so it is useful
to define the constant quantity $\xi_i \equiv \mu_i/T_i$.
Then the neutrino and antineutrino number densities
are functions of $\xi_i$:
\begin{equation}
\label{nnu1}
n_{\nu_i} = {1\over 2 \pi^2} T_\nu^3 \int_0^\infty {x^2 dx \over 1+
\exp(x - \xi_i)},
\end{equation}
and
\begin{equation}
\label{nnu2}
n_{\bar \nu_i} = {1\over 2 \pi^2} T_{\bar \nu}^3 \int_0^\infty {x^2 dx\over 1+
\exp(x + \xi_i)},
\end{equation}
and
the total
energy density of the neutrinos and antineutrinos is
\begin{equation}
\label{rhonu}
\rho = {1\over 2 \pi^2} T_\nu^4 \int_0^\infty {x^3 dx\over 1+
\exp(x - \xi_i)}
+ {1\over 2 \pi^2} T_{\bar \nu}^4 \int_0^\infty {x^3 dx\over 1+
\exp(x + \xi_i)}.
\end{equation}
Electron neutrino degeneracy changes the $n \leftrightarrow p$
weak rates through the number densities
given in equations (\ref{nnu1}) and (\ref{nnu2}),
while the change in the expansion
rate due to the altered energy density in equation (\ref{rhonu}) affects BBN
for degeneracy of any of the three types of neutrinos.
(See Ref. \cite{wagoner}
for a more detailed discussion).

Now consider the effect of inhomogeneities in the neutrino chemical
potential.  As noted in Ref. \cite{whitmire}, neutrino free-streaming
will erase any fluctuations on length scales smaller than the horizon at
any given time.  Thus, in order for inhomogeneities to affect BBN,
they must be non-negligible on scales larger than the horizon
scale at $n \leftrightarrow p$ freeze-out, which corresponds to a comoving
scale $\sim 100$ pc today.  On the other hand, if the neutrino chemical potential
is inhomogeneous on scales larger than the element diffusion scale,
estimated in Ref. \cite{whitmire} to correspond to a comoving length
$\sim 1$ Mpc, then the result will be an inhomogeneous distribution of
observed element abundances today (the possibility considered in
Ref. \cite{dolgov2}).

To make any further progress, we need a specific distribution $f(\xi_i)$
for neutrino chemical potentials.  In analogy with the distribution of
primordial
density perturbations (and in accordance with the central
limit theorem) we take this distribution to be a multivariate
Gaussian.  Such a distribution is entirely characterized by the power
spectrum of fluctuations, $P(k)$.  For a power spectrum of
the form $P(k) \propto k^n$
the rms fluctuation $\sigma_\xi$ on a given length scale $\lambda$
is given by
\begin{equation}
\sigma_\xi \propto \lambda^{-(n+3)/2}
\end{equation}
We wish to consider only cases
for which the presently-observed element distribution (determined
by $\sigma_\xi$ at a comoving scale of $\sim 1$ Mpc) is homogenous,
while the distribution is highly inhomogenous on the horizon scale
at nucleosynthesis (a comoving scale of $\sim 100$ pc).
Since our two length scales of interest differ by a factor of $10^4$,
this condition can be satisfied for any $k^n$ power spectrum with $n > -3$.
For instance, for a white-noise power spectrum, $n=0$, a value
of $\sigma_\xi = 1$ at the BBN horizon scale corresponds to
$\sigma_\xi = 10^{-6}$ at the element diffusion scale.

Given these conditions, it is a good approximation to assume that
BBN takes place in separate horizon volumes, with
the value of $\xi$ taken to be homogeneous within each volume.
At late times, the elements produced within each volume
mix uniformly to produce the observed element abundances
today.

We make the additional assumption that the neutrino fluctuations
are isocurvature, so that the total fluctuation in energy density
is zero, even when the chemical potential is inhomogeneous.
This implies that the overdensity in the degenerate neutrinos is
compensated by an underdensity in some other component.
In Ref. \cite{dolgov2}, for example, the degeneracies in each
of the three neutrinos are arranged so that the total density remains
uniform.  In Ref. \cite{dibari}, the compensation is produced by
a sterile neutrino.  Such models have the advantage that they produce
no additional inhomogeneities in the cosmic microwave background
as long as the compensating energy density does not include photons or baryons.
(Note that this is not the assumption made in Ref. \cite{whitmire}).
We also assume for simplicity that $\eta$ remains uniform
in the presence of an inhomogeneous lepton distribution.
With this set of assumptions, the only neutrino for which inhomogeneities
in the chemical potential are important for BBN
is the electron neutrino; the effect of the other neutrino
chemical potentials
is to alter the total energy density, which is now assumed to be homogeneous.

It has recently been noted that if the large mixing angle solution of
the solar neutrino problem is correct, then neutrino
flavor oscillations will cause the neutrino chemical potentials
to equilibrate prior to Big Bang nucleosynthesis
\cite{lunardini,dolgov3,abazajian}.  In our inhomogeneous model,
the effect of this equilibration
would
depend on the compensation mechanism for the inhomogeneities.  In models
in which the fluctuations in
the electron neutrino chemical potential are compensated by fluctuations
in the chemical potentials of the $\mu$ and $\tau$ neutrinos, the effect
of such flavor oscillations would be to erase any spatial fluctuations in
the chemical potentials.  In models where the electron neutrino chemical
potential
fluctuations are compensated in some other way, the chemical potentials
of all three species would be equal at any point in space, but
the spatial fluctuations would be preserved.  Any large $\Delta N_\nu$
in this case would have to be due to some other form of energy beyond
the standard three neutrinos.

\section{Calculations and Discussion}

The model described in the previous section
can be completely specified by two parameters,
the (inhomogeneous) electron neutrino degeneracy parameter, $\xi_e$,
and the additional (homogeneous) energy density due to the
degeneracy of all three neutrinos plus any additional
relativistic component.  We parametrize the latter in terms
of $\Delta N_\nu$,
the effective number of additional neutrinos.  This second parameter
hides our ignorance about the compensation mechanism and 
about the degeneracies among the other two
types of neutrinos.
In our simulation, we take $\xi_e$
to be homogeneous within
a given horizon volume during nucleosynthesis.
Different horizon volumes may have different values
of $\xi_e$, which are given by the distribution function $f(\xi_e)$,
i.e, the probability that a given horizon
volume has a value of $\xi_e$ between $\xi_e$
and $\xi_e + d\xi_e$.  (Since we are considering only inhomogeneities
in electron neutrinos, we now drop the $e$ subscript).
We
take $f(\xi)$ to have a Gaussian distribution with mean $\bar \xi$
and rms fluctuation $\sigma_\xi$:
\begin{equation}
f(\xi) = {1 \over \sqrt {2 \pi}\sigma_\xi} \exp[- (\xi - \bar \xi)^2/
2\sigma_\xi^2]
\end{equation}
Then the final primordial element abundances, for a fixed value of
$\eta$ and $\Delta N_\nu$, will be functions of $\bar \xi$ and $\sigma_\xi$;
we can write, for a given
nuclide $A$,
\begin{equation}
\label{XA}
\bar X_A = \int_{-\infty}^\infty X_A(\xi)f(\xi) d\xi,
\end{equation}
where $X_A(\xi)$ is the mass fraction of $A$ as
a function of $\xi$, and
$\bar X_A$ is the mass fraction of $A$ averaged
over all space; after the matter is thoroughly mixed, $\bar X_A$
will be the final primordial mass fraction.

A full treatment for all possible values of $\eta$, $\Delta N_\nu$,
$\bar \xi$, and $\sigma_\xi$ is impractical.  We have chosen to
concentrate on variations in the latter two quantities,
since we are most
interested in the effects of inhomogeneities in the chemical potential.
Because of the large number of
free parameters and the difficulty of
exhaustively searching all of parameter space, our goal
is to discern any general results which
are independent of $\eta$ and $\Delta N_\nu$.

There are now strong limits on $\eta$ from the cosmic microwave
background alone, independent of BBN.  We examine two
extreme values for $\eta$:
$\eta = 4 \times 10^{-10}$ and $\eta = 1 \times 10^{-9}$;
these represent very conservative lower and upper bounds
on $\eta$ from the CMB in models with non-zero
neutrino degeneracy \cite{Kneller}.  For $\Delta N_\nu$,
we consider $\Delta N_\nu = 0$ and 5.  Note that the
first of these is only possible if the extra energy density
in the degenerate
electron neutrinos is compensated by a decrease in the
energy density in some other relativistic
component.  
For each of these cases, we calculate the abundances of $^4$He,
D, and $^7$Li as a function of $\sigma_\xi$ for
$\bar \xi = -1$ to $+1$ in steps of 0.5.  Our results
are displayed in Figures 1-3.  In each of these figures, we also show
observational limits on the primordial element abundances from Ref.
\cite{olive}:
$2.9 \times 10^{-5} < ({\rm D/H}) < 4.0 \times 10^{-5}$, $0.228 < Y_P < 0.248$,
and $-9.9 < \log(^7{\rm Li/H}) < -9.7$.

The general behavior of the element abundances in Figs. 1-3 is
very clear.  As expected, for $\sigma_\xi \ll \bar \xi$,
the abundances of deuterium, $^4$He, and $^7$Li are
unchanged from their values in the corresponding homogeneous
model with the same value of $\bar \xi$.  At the opposite limit,
when $\sigma_\xi \gg \bar \xi$, the models all converge
to a single limiting value; again, this is what one would naively
expect.  What is interesting is that, with a few exceptions, 
the introduction of a Gaussian distribution
of values for $\xi$ results in an {\it increase} in the abundance of each
element
relative to the corresponding homogeneous model with the same value of
$\bar \xi$.  The only exceptions occur for $^4$He with
negative values of $\bar \xi$, (for which $Y_P$ is far too large
to be physically reasonable), and some of the $^7$Li curves, for which
there is a tiny decrease in the $^7$Li abundance over
a short range of $\sigma_\xi$ values.

This result may seem surprising,
but it is a simple
consequence of the behavior of $X_A(\xi)$.  In particular,
if $X_A(\xi)$ is a convex function ($X_A^{\prime \prime}(\xi) > 0$),
then Jensen's inequality \cite{feller} gives
\begin{equation}
\label{jensen}
\int_{-\infty}^\infty X_A(\xi)f(\xi) d\xi > X_A(\bar \xi).
\end{equation}
We find, for example, for $\Delta N_\nu = 0$, and
both values of $\eta$, that our
$X(\xi)$ curves are all convex in the range $-2 < \xi < 2$,
with the exception of $^4$He at $\xi < -1$, and $^7$Li
with $\eta = 1 \times 10^{-9}$.  These are precisely
the regimes for which we observe equation (\ref{jensen}) to fail.
Of course, none of the $X_A(\xi)$ curves is convex
for all values of $\xi$; the practical condition for equation
(\ref{jensen}) to hold is that the $X(\xi)$ curves be convex as long
as $f(\xi)$ is non-negligible.

\onecolumn

\begin{figure}
\begin{center}
\epsfysize=8.0truein
\epsfbox{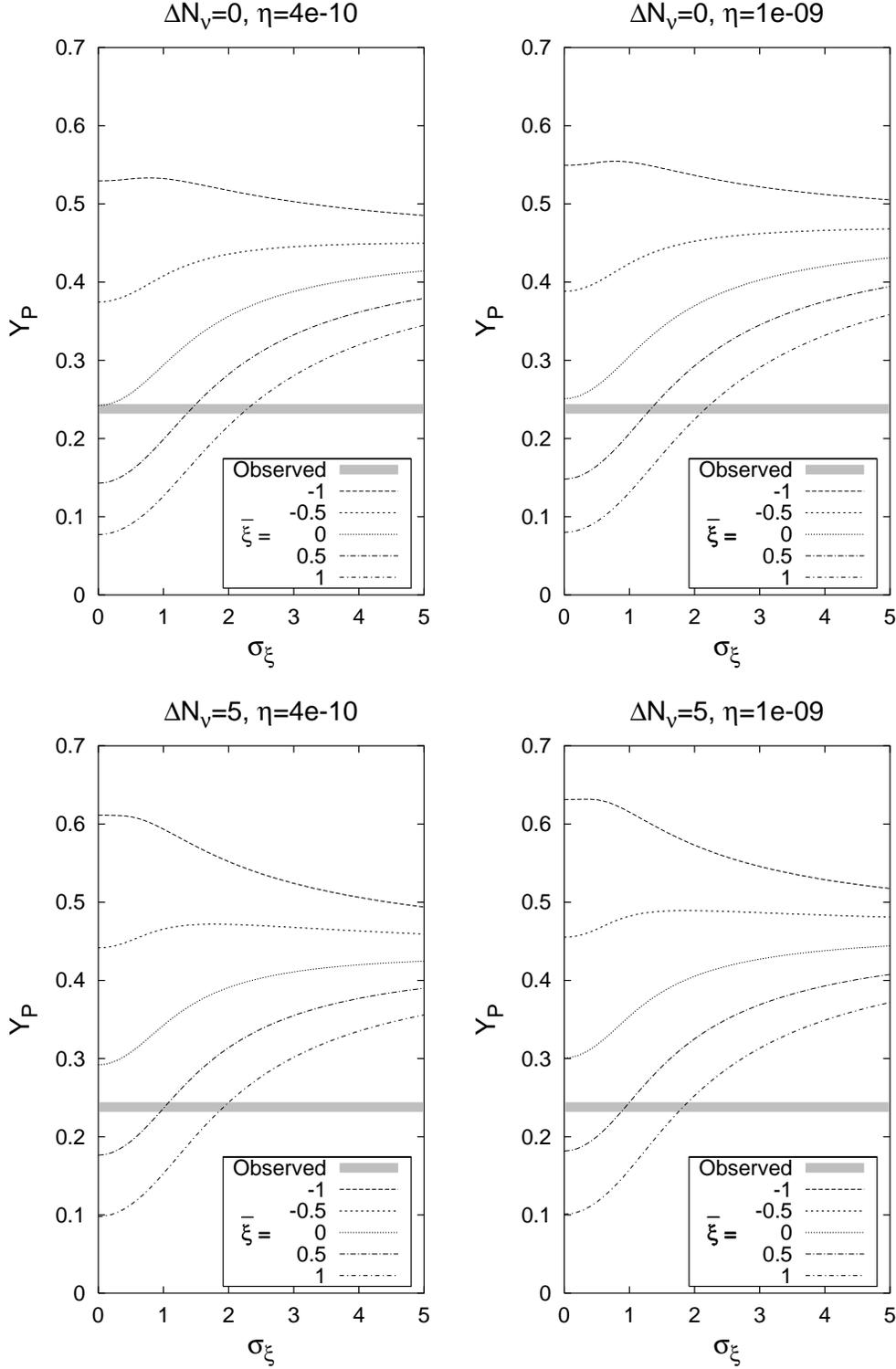}
\vspace*{0.5truecm}
\end{center}
\vspace*{-0.3truecm}
\caption{The primordial $^4$He mass fraction, $Y_P$, as a function
of the rms fluctuation in the electron chemical potential $\sigma_\xi$,
for the indicated value of the mean electron neutrino chemical
potential $\bar \xi$.  Each figure corresponds to the indicated value of
the baryon-photon ratio $\eta$ and
the effective number of extra neutrinos $\Delta N_\nu$.  The gray
shaded region gives observational limits on $Y_P$
from Ref. [19].}
\label{fig1}
\end{figure}

\begin{figure}
\begin{center}
\epsfysize=8.0truein
\epsfbox{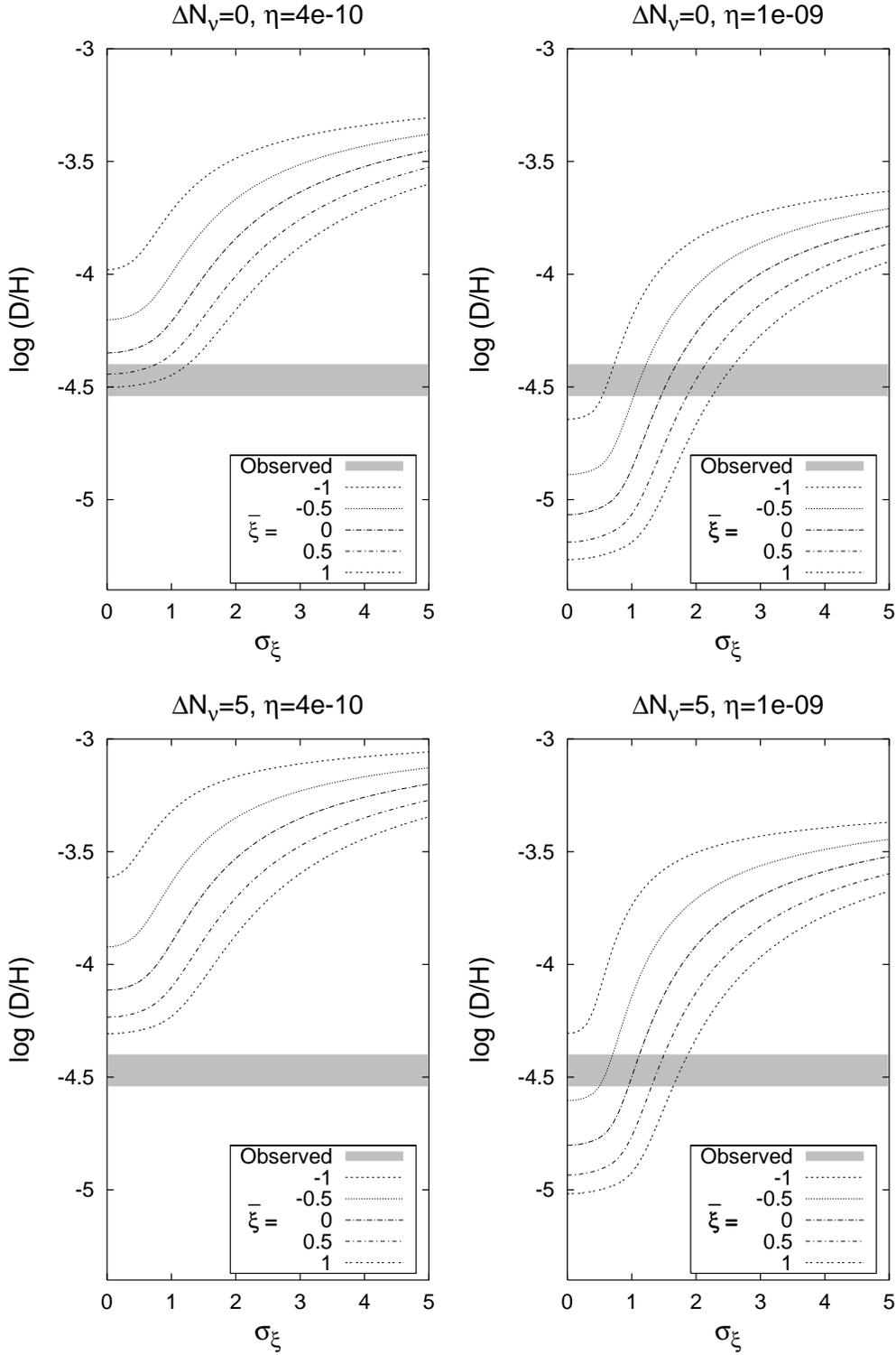}
\end{center}
\baselineskip 7pt
\caption{As Fig. 1, for the primordial ratio of deuterium to hydrogen,
(D/H).}
\label{fig2}
\end{figure}

\begin{figure}
\begin{center}
\epsfysize=8.0truein
\epsfbox{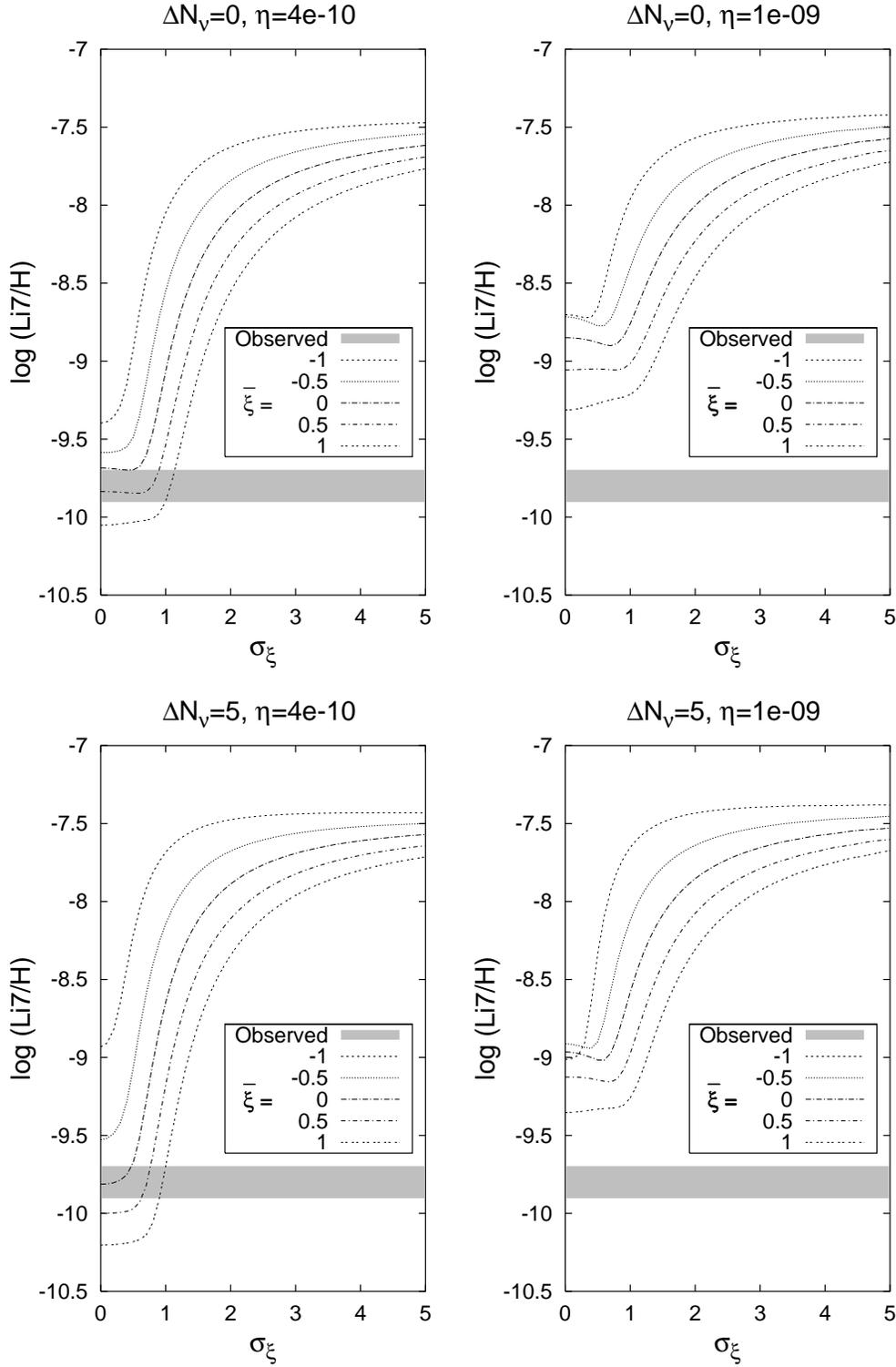}
\end{center}
\caption{As Fig. 1, for the primordial ratio of $^7$Li to hydrogen,
($^7$Li/H).}
\label{fig3}
\end{figure}

\twocolumn

This simple behavior allows us to draw some useful general
conclusions.  In models in which $\xi$ and $\Delta N_\nu$ are allowed
to vary freely, if we fix $\xi$ and
trace out the allowed region in the $\eta$, $\Delta N_\nu$ plane,
then the upper and lower bounds on $\eta$ are set
primarily by the upper observational bound on $^7$Li and the upper
observational limit on D, respectively, with the $^4$He limits
serving primarily to set the bounds on $\Delta N_\nu$ \cite{Kneller}.  
However, our results indicate that the general effect of
going from a homogeneous to an inhomogeneous distribution in
$\xi$ is to {\it increase} both the deuterium and the $^7$Li
abundances.  (Again, we note a slight
decrease in $^7$Li over a small range in $\sigma_\xi$, but
this is a tiny effect).  Hence, the net effect of introducing this inhomogeneity
will be to {\it decrease} the allowed range for $\eta$, in comparison
with the corresponding homogeneous model.  This is a rare example
in the study of BBN in which the introduction
of an extra degree of freedom does nothing to increase the allowed range
for $\eta$.
Instead, the effect of adding a Gaussian
distribution of values for $\xi$ is to decrease the allowed
range for $\eta$.

Although we have assumed a Gaussian distribution for $\xi$, our
results are much more general.
In particular, as long as our distribution $f(\xi)$ is
negligible over
the range of values of $\xi$ for which
$X_A(\xi)$ is not a convex function, we expect equation
(\ref{jensen}) to hold.
This would apply, for example, to a top hat distribution with
the same values of $\sigma_\xi$ as those examined here.
Moreover, the distribution $f(\xi)$ need not even be symmetric
for our results to apply.

Our results contrast with those of Ref. \cite{whitmire},
which found an expanded upper limit on $\eta$ in models
with inhomogeneous $\xi$.
The reason for this difference is that the models examined
in Ref. \cite{whitmire} allowed for an arbitrary distribution in $\xi$,
and large increases in $\eta$ occurred for bizarre
distributions in $\xi$.  In particular, the distributions in
Ref. \cite{whitmire}
sampled extreme values for $\xi$, outside the range for which
all of the $X_A(\xi)$ functions are convex.

\acknowledgements

We thank N. Bell, S. Pastor, G. Steigman for helpful comments on the manuscript.
S.D.S. was supported at Ohio State under the NSF
Research Experience for Undergraduates (REU) program
(PHY-9912037).
R.J.S. is supported by the Department of Energy
(DE-FG02-91ER40690).

%
%

\end{document}